\documentclass[aps,prb,showpacs]{revtex4}

\usepackage{graphicx}
\DeclareGraphicsExtensions{.eps}

\begin{document}

\title{Micromagnetic simulations of the magnetization precession induced by 
a spin polarized current in a point contact geometry}
\author{D.V.~Berkov, N.L.~Gorn}
\affiliation{Innovent e.V., Pr\"uessingstr. 27B, D-07745, Jena, Germany}

\date{\today}

\begin{abstract}

This paper is devoted to numerical simulations of the magnetization dynamics driven
by a spin-polarized current in extended ferromagnetic multilayers when a point-contact 
setup is used. We present (i) detailed analysis of methodological problems 
arising by such simulations and (ii) physical results obtained on a system
similar to that studied in Rippard et al., Phys. Rev. Lett., {\bf 92}, 027201 (2004).
We demonstrate that the usage of a standard Slonczewski formalism for the phenomenological
treatment of a spin-induced torque leads to a qualitative disagreement between simulation 
results and experimental observations and discuss possible reasons for this discrepancy.
 
\end{abstract}


\maketitle

\section{Introduction}

Nearly a decade after prediction \cite{SpInjPred} and subsequent experimental 
discovery \cite{SpInjDiscov} of magnetic excitation and magnetization switching
induced by a spin-polarized current (SPC) in a thin magnetic film high-quality experiments
providing quantitative information concerning the corresponding magnetization 
dynamics have been performed (see, e.g., Ref. \onlinecite{Kiselev2003,Kiselev2004,
Rippard2003,Rippard2004a,Rippard2004b}). For this reason the important problem of the 
{\it quantitative} verification of existing theoretical models for the spin-transfer
phenomena and SPC-induced magnetization dynamics \cite{SpInjTheory} can be addressed.
Corresponding research clearly requires full-scale micromagnetic simulations, because
single-domain (macrospin) approximation \cite{Sun2000} by its definition can not 
incorporate important effects resulting from the inhomogeneity of the magnetization 
states. Such strongly non-uniform magnetization configurations during the SPC-induced 
precession are predicted already for thin film elements of very small sizes 
\cite{BerkovChaos2005} ($\sim 30$ nm), which are less than typical dimensions of any 
actually used experimental samples.

Micromagnetic simulations performed up to now deal only with the experiments performed
in the so called columnar geometry \cite{SimulColumnGeom}, where an electric current flows
through a multilayer magnetic element with very small lateral sizes $\sim 10^2$ nm. Such
simulations were able to reproduce many important features of the experimental data
obtained on columnar structures, in particular, the existence of regular and quasichaotic 
oscillation regimes, dependence of the oscillation frequency on the current strength etc. 
(however, we mention that even the most sophisticated model failed to reproduce experimental 
data quantitatively \cite{BerkovEllipse2005}). In contrast to this situation, to the best
of our knowledge, simulations of the point-contact experiments \cite{Rippard2004a,Rippard2004b},
which show some important magnetization dynamics features different from those observed on
columnar structures, have not been carried out.

In this paper we present such full-scale micromagnetic simulations of the SPC-induced 
magnetization dynamics in the point-contact setup. The paper is organized as follows: in
Sec. \ref{sec:NumSimMeth} we discuss methodological problems of such simulations which
are specific for the geometry under study. Next (Sec. \ref{sec:NumSimRes}) we present 
our results, starting from the analysis of the magnetization dynamics without the
current-induced magnetic field (which enables a much more transparent presentation
of several important effects). Brief comparison with the experimental data is
performed in the last section.

\section{Methodological problems of numerical simulations in the point contact setup}
\label{sec:NumSimMeth}

Numerical simulations were carried out with our software package MicroMagus 
\cite{MicroMagus}.This package uses the modified Bulirsch-Stoer algorithm with the adaptive 
step-size control for integrating the Landau-Lifshitz-Gilbert (LLG) equation 
to calculate the time evolution of the system magnetization configuration. 
For the additional torque created by the spin polarized current (SPC) we have assumed 
the symmetric Slonczewski form 
$\Gamma = (a_J/M_S) \cdot [{\bf M} \times [{\bf M} \times {\bf S}]]$ (${\bf S}$ is 
the spin polarization direction) in order to study the SPC-induced dynamics in the simplest 
possible approximation. In the trilayer system under study (see below) the spin torque 
was assumed to act on the on the magnetization of the 'free' layer only. The site
dependence of the spin torque magnitude $a_J({\rm r})$ (in the first approximation
confined within the point contact area) is discussed below. 

To enable a comparison with the experimental results reported in the most advanced
quantitative studies of magnetization oscillations in the point contact geometry 
\cite{Rippard2004a,Rippard2004b}, we have chosen the system parameters as close 
as possible to those reported in Ref. \onlinecite{Rippard2004a}. We have simulated 
a trilayer system consisting of two magnetic layers and an interlayer non-magnetic spacer:
the lower 'fixed' layer (underlayer) with the thickness $h_1 = 10$ nm and magnetic parameters typical
for Co$_{90}$Fe$_{10}$ (saturation magnetization $M_S = 1500$ G, exchange constant 
$A = 2 \times 10^{-6}$ erg/cm); the upper (thin) Permalloy-like magnetic layer with
the thickness $h_2 = 5$ nm and $M_S = 640$ G (as measured in Ref. \onlinecite{Rippard2004a}) 
and $A = 1 \times 10^{-6}$ erg/cm (standard value for Py \cite{Doering1966}); the spacer 
thickness was set to $h_{sp}=5$ nm as for the Cu spacer used in 
Ref. \onlinecite{Rippard2004a}. All results presented below were obtained for an external
field $H_0 = 1000$ Oe. Coordinate axes $0x$ and $0z$ lie in the film plane, with the 
$x$-axis directed along the external field.

The 'fixed' layer thickness $h_1 = 10$ nm was taken less than the experimental
value \cite{Rippard2004a} $h_1^{\rm exp} = 20$ nm, because proper simulations of such 
relatively thick layers require not only their in-plane discretization, but also the 
subdivision into sublayers, which would lead in this case to prohibitively large 
computational times; the influence of underlayer thickness on the magnetization 
dynamics will be discussed elsewhere. 

We also did not study the effect of the polycrystalline structure of Co$_{90}$Fe$_{10}$ 
underlayer, although the magnetocrystalline anisotropy of this material is not negligible 
(cubic anisotropy with $K_1 \approx 5.6 \times 10^5$ erg/cm$^3$ was reported in 
Ref. \onlinecite{CoFeAnis}). The magnetization dynamics of the system under study 
turned out to be very complicated already for ideal layers (without taking into 
account their polycrystalline structure and corresponding random anisotropy), so we 
have postponed the study of the random magnetocrystalline anisotropy effects.

In contrast to micromagnetic study of spin injection effects observed in the columnar 
structures \cite{Kiselev2003,Kiselev2004}, where standard micromagnetic methods can be 
applied \cite{SimulColumnGeom}, simulations of the point-contact experiments encounter 
serious methodological problems. Here we discuss two of them: (i) artificial interference 
effects occurring both for open and periodic boundary conditions and (ii) artificial 
short-wave magnetization oscillations arising by the usage of a sharp cut-off of charge 
and spin currents.

{\it Artificial interference effects}. This problem occurs due to the combination of two
circumstances: first, for practical purposes magnetic materials with low dissipation rate
$\lambda \sim 0.01 - 0.02$ are used (to ensure a small linewidth of the oscillation spectrum
and a reasonably low excitation threshold), and second, a lateral size of a system available
for simulations is much smaller than that used experimentally. In real experiments the 
lateral size of a multilayer is about 10 mkm \cite{Rippard2004a}, which is definitely 
far above the value accessible for numerical simulations, especially taking into account 
that dynamic simulations require a much finer mesh than quasistatic ones.
In particular, we have found out that to obtain a mesh-independent results for a system
with geometric and magnetic parameters given above, a lateral discretization as fine as 
$\Delta x \times \Delta z = 2.5 \times 2.5$ nm$^2$ is necessary, which limits the
simulated lateral system size to $\sim 1$ mkm$^2$ (which means more $\sim 10^5$ cells 
per layer). In a layer made of a material with low dissipation (see above) the decay
length of the spin wave with the wavevector corresponding to the inverse size of the point 
contact has the same order of magnitude as the simulated area size. 

When open boundary conditions (OBC) are used (i.e., a finite size element is simulated), 
this means that the wave emitted by the point contact propagates across the whole element, 
is reflected at its free borders and returns to the contact location. The strong 
interference of this reflected wave with primary (SPC-induced) magnetization oscillations 
leads to unphysical artifacts, especially taking into account that both waves have the same 
frequency. Another problem arising by using OBC is a complicate pattern of the wave 
reflection due to the inhomogeneous magnetization configuration on the element edges.

For periodic boundary condition (PBC) the magnetization configuration at the simulated area
borders is homogeneous (which is the main reason to use PBC), but the primary wave
also propagates across the whole simulated area and due to PBC enters this area from the
opposite side, causing the same undesirable interference effects. To eliminate this effects,
a method to absorb the wave near the simulation area borders, not affecting the low dissipation
at and near the point contact, is required.

To ensure such an absorption, we have embedded in our code an artificial site dependence of
the dissipation coefficient $\lambda({\bf r})$. The function describing this  
dependence should fulfil several conditions: (i) the dissipation within and nearby 
the point contact area should remain equal to its physical value $\lambda_0$ to preserve 
the dynamic properties of the system under study, (ii) the dissipation coefficient far from
the point contact (near the simulation area borders) should be large enough to ensure
the wave energy absorption, (iii) spatial variation $\lambda({\bf r})$ should be smooth
enough to prevent the wave reflection from the border between the areas of small and 
relatively large $\lambda$ (due to the abrupt changes of the media properties). 
Following these requirements, we have adopted a site-dependent dissipation 
$\lambda({\bf r})$ in the form 
\begin{equation}
\label{LamSiteDep}
\lambda({\bf r}) = \lambda_0 + \Delta\lambda 
\left(
1+ \tanh {r - R_0 \over \sigma_{\lambda}}
\right)
\end{equation}

Here it is assumed that the point contact is located at the coordinate origin. The
function (\ref{LamSiteDep}) provides a gradual increase of the dissipation parameter 
above the (small) basic value $\lambda_0$ which starts at the distance
$\approx (R_0 - \sigma_{\lambda})$ from the point contact and occurs smoothly within 
a ring of the width $\approx 2\sigma_{\lambda}$. The maximal dissipation value reached 
outside of this ring is $\lambda_{\rm max} = \lambda_0 + \Delta\lambda$. We have found out 
that for the system size $L_x \times L_z = 1 \times 1$ mkm$^2$ and basis dissipation 
values in the range $\lambda_0 = 0.01 - 0.04$ the introduction of the additional 
dissipation (\ref{LamSiteDep}) with $R_0 = 300$ nm, $\sigma_{\lambda} = 40 nm$ and
$\Delta\lambda = 0.1$ ensured the wave absorption at the simulation area borders,
not changing the magnetization dynamics within and around the point contact area.
For simulations considered here we have used the basis dissipation 
value $\lambda_0 = 0.02$.

{\it Spurious magnetization oscillations caused by a sharp spatial cut-off of 
a point-contact current}. By simulations of the columnar geometry  
the current is usually assumed to be distributed homogeneously within the layer plane 
of a nanoelement, which does not lead to any methodical problems because the magnetization 
is also present only inside the area where the current flows. In contrast to this 
simple situation, by simulations of a point-contact setup the naive usage of 
the step-like current density in the form $j(r \le D/2) = j_0$ and $j(r > D/2) = 0$ 
($r$ being the distance from the contact center, $D$ - the contact diameter) 
lead to the development of artificial magnetization oscillations with the smallest 
wavelength supported by the given lattice. The reason for these oscillations 
is the non-physical abrupt change of the current density $j({\bf r})$ at $r = D/2$, 
so that its spatial Fourier image and the Fourier image of the current-induced 
magnetic field (the Oersted field) $H_{\rm Oe}({\bf k})$ exhibits large tails 
up to the highest values of the wavevectors available for the simulated discrete 
system. This leads to artificial instabilities for these wavevectors, resulting 
in the appearance of corresponding magnetization oscillations.

In order to avoid this problem, we have smoothed the spatial distribution
of $H_{\rm Oe}({\bf r})$ obtained in the approximation of a sharp electric current 
cut-off convolving it with the Gaussian kernel $\exp(-r^2/2\sigma_H^2)$. A physically
meaningful choice of the smoothing parameter $\sigma_H$ would require a reliable 
information about the lateral diffusion of the electric current carriers to calculate
the actual spatial distribution of the current density. Lacking such knowledge,
we have simply adopted the minimal value $\sigma_H = 2 \Delta x$ (two times larger than
the mesh size) which was sufficient to eliminate the artificial oscillations mentioned
above. Further increment of this parameter within a reasonable range (up to 
$\sigma_H = (4 - 5) \cdot \Delta x$) had only a minor influence on physical results.

A similar problem is caused by the sharp spatial cut-off of the {\it spin} current
density represented by the amplitude $a_J({\bf r})$ of the SPC-induced torque
$\Gamma = (a_J/M_S) \cdot [{\bf M} \times [{\bf M} \times {\bf S}]]$,
although spatial oscillations caused by this cut-off are weaker than discussed
in the previous paragraph. Nevertheless, the same kind of smoothing is also required 
to solve this problem. The smoothing parameter of the corresponding kernel 
$\exp(-r^2/2\sigma_S^2)$ is directly related to the spin diffusion length and could 
in principle be computed from the corresponding theory. In this study, however, we
have also simply used the value $\sigma_S = 2 \Delta x$ for the same reasons as
explained above for $\sigma_H$. At this point we would like to emphasize, that - in 
contrast to the Oersted field smoothing parameter - the $\sigma_S$-value significantly
influences the system behaviour; in particular, the threshold value of $a_J$ for the
onset of steady-state microwave oscillations substantially depends on $\sigma_S$.
For this reason the problem of calculating the actual distribution of a {\it spin} 
current in the point contact geometry deserves spatial attention.

All results presented below were obtained employing the site-dependent damping 
(\ref{LamSiteDep}) and smoothing of the Oersted field and spin current distribution 
with parameters given above. The in-plane discretization of both magnetic layers with 
the mesh size $\Delta x \times \Delta z = 2.5 \times 2.5$ nm$^2$ and the full size of 
the simulation area $L_x \times L_z = 1 \times 1$ mkm$^2$ (with PBC) were used. 
The point contact diameter was set to $D = 40$ nm.

\section{Numerical simulations: Results and discussion}
\label{sec:NumSimRes}

In this paper we discuss only simulation results obtained without taking into account
the effect of thermal fluctuations ($T = 0$). Even without these effects the system 
demonstrates very complicated dynamics, which should be understood before thermal 
fluctuations are included into consideration.

{\it Dynamics without taking into account the Oersted field}. We start with the analysis
of the magnetization dynamics when the influence of the Oersted field is neglected. We 
point out already here that in the experimental situation \cite{Rippard2004a} 
the current-induced magnetic field is comparable with the externally applied field: 
For the total current $I \sim 4$ mA flowing through the point contact with the diameter 
$D \approx 40$ nm the maximal value of the Oersted field is 
$H_{\rm Oe}^{max} \sim 400$ Oe, whereas the external field used in 
Ref. \onlinecite{Rippard2004a} to present most detailed results is $H_{\rm ext} = 1000$ Oe.
For this reason we do not expect the approximation when $H_{\rm Oe}$ is neglected
to be quantitatively correct, but neglecting the Oersted field simplifies the magnetization
dynamics of the system under study, preserving most of its qualitative features,
which can be thus demonstrated more clearly.

\begin{figure}[tbph] 
\centering
{\includegraphics
[scale=0.8, bb=5cm -2cm 15cm 23cm]
{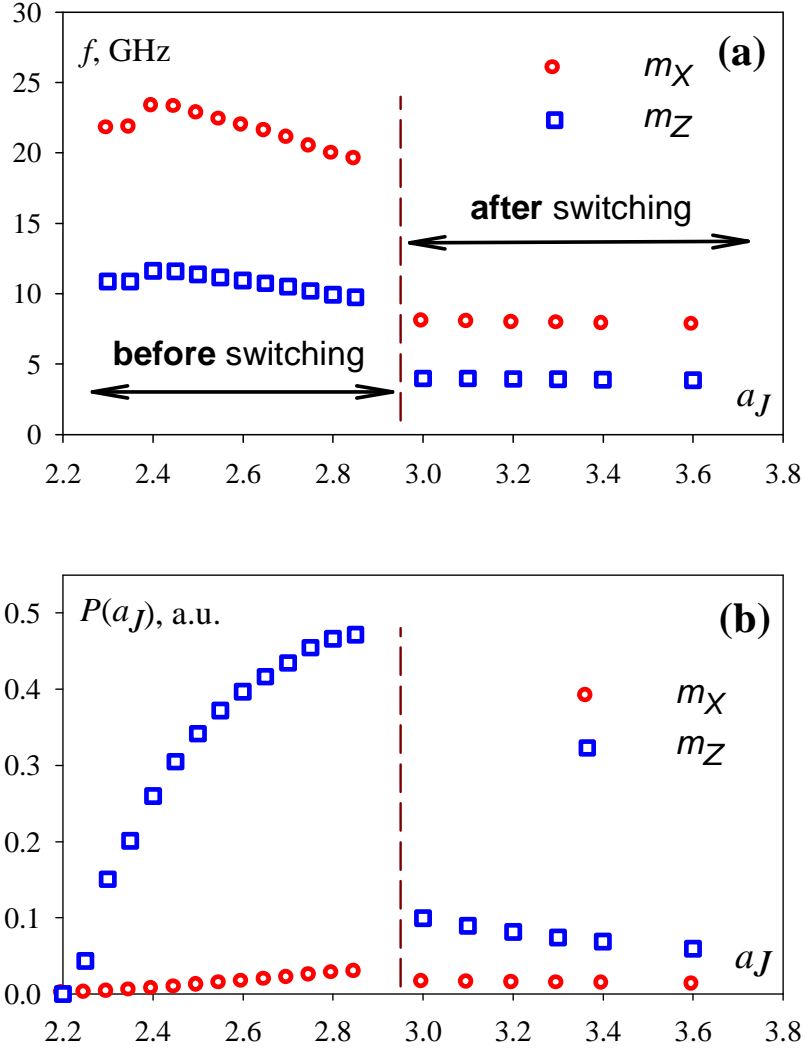}}
\caption
{
Dependencies of the oscillation frequency (a) and the total oscillation power (b) of 
$m_x^{\rm av}$ (circles) and $m_z^{\rm av}$ (squares) magnetization projections (averaged 
over the point contact area) on the spin current strength given by the Slonczewski torque
amplitude $a_J$. The $a_J$-value where the magnetization within the contact area switches 
to the direction opposite to the external field, is marked by the vertical dashed line.
}
\label{fig:TwoPrecRegimes_noOeField} 
\end{figure}

The major feature of the simulated magnetization dynamics in the point contact geometry 
is the existence of {\it two} current regions where the steady-state precession of the
magnetization within and nearby the point-contact area exists 
(Fig. \ref{fig:TwoPrecRegimes_noOeField}).

In the first current region - {\it before} the magnetization in the point contact area is 
switched under the SPC influence, i.e., when the magnetization is (on average) still 
directed along the external field, the magnetization dynamics is relatively simple.
Magnetization configuration of the thin (upper) layer within the point-contact area remains
roughly collinear. Spin waves emitted from the area under the contact are smooth and have 
a simple elliptical wavefront (Fig. \ref{fig:TwoPrecRegimes_noOeField}). The limit oscillation 
cycle of the magnetization ${\rm m}^{\rm av}$ averaged over this area represents a slightly bended ellipse
(Fig. \ref{fig:PrecBeforeSw_noOeField}), time dependencies of the magnetization components 
are nearly ideal harmonic functions, so that oscillation power spectra consist of a single 
and very narrow peak. As usual for such in-plane oscillations, the frequency of 
$m_x^{\rm av}$-oscillations (the component along the external field $H_{\rm ext}$) 
is twice the $m_z^{\rm av}$-frequency (the in-plane component perpendicular to $H_{\rm ext}$),
because $m_x^{\rm av}$ goes back and force twice during a single oscillation cycle.
The oscillation frequency decreases monotonically with increasing current ($a_J$ in our
formalism), which is mainly due to the increase of the oscillation amplitude (longer
limit cycle) with the current strength. The oscillation power sharply increases when
the current exceeds the threshold for the oscillation onset and then growth smoothly
until the magnetization under the point contact area is switched by the SPC.

It turns out, however, that in the model simulated here a steady-state precession exists
also {\it after} the point contact area switching caused by spin injection. By the 
transition to this second regime we observe a large frequency jump - for the system 
parameters used in this study the frequency drops down from $f_{\rm bef} \approx 9.8$ GHz
to $f_{\rm aft} \approx 4.0$ GHz and then remains almost current-independent. Although
the limit cycles have in this regime a more complicated form and the time dependencies 
of the magnetization components considerably differ from ideal sinusoids, spectral 
lines are still so narrow, that within the physical time corresponding to the longest 
simulation run performed ($\approx 20$ ns) their width could not be resolved 
($\Delta f \le 50$ MHz). Oscillation amplitude slowly decreases with current leading
to the corresponding decrease of the oscillation power.

Although spectral lines in this second precession mode remain quite narrow, the magnetization
configurations appearing during the precession are extremely complicated even when the
Oersted field is neglected. First of all we note that the precession frequency is below
the frequency of the homogeneous FMR mode for the layer under consideration 
($f_0 = (\gamma/2\pi) \cdot (H_0(H_0 + 4\pi M_S))^{1/2} \approx 8.4$ GHz. For this reason
the 'normal' circular (elliptical) wave can not exist in this regime, so that the energy is emitted in
form of the soliton-like wave packages, as shown in Fig. \ref{fig:PrecAfterSw_noOeField}.
The magnetization within and nearby the contact area itself forms vortex/antivortex pairs
(the latter state is also sometimes called a cross-like configuration), which creation and
annihilation is the basic feature of the magnetization dynamics in this regime; a typical
example of such a structure is shown in Fig. \ref{fig:ArrowMagnConf_noOeField}. Detailed
analysis of these challenging structures will be performed elsewhere. 

{\it Dynamics with the Oersted field included}. Inclusion of the Oersted field requires
the establishing of the relation between the current strength used in the actual experiment
and the parameter $a_J$ used in simulations to set the amplitude of the SPC-induced torque.
Lacking the exact microscopic theory which could provide such a relation, we have used 
the same procedure as in Ref. \onlinecite{BerkovEllipse2005}, i.e., we assumed that 
the oscillation onset threshold $a_J = 2.2$ corresponds to the minimal current 
$I_{\min} \approx 4$ mA where the magnetization precession is observed experimentally.

The current-induced magnetic field $H_{\rm Oe}$, being strongly inhomogeneous, results 
in a much more complicated magnetization dynamics than in the absence of $H_{\rm Oe}$. 
The most obvious change is the appearance of the wave asymmetry in the steady-state
precession regime before switching: in that half of the layer where the Oersted field 
$H_{\rm Oe}$ is directed opposite to the external field $H_0$ (thus partly compensating 
it), the wave amplitude is significantly larger than in the other half of the film (where 
the external 
field is enhanced by $H_{\rm Oe}$). The inhomogeneity of $H_{\rm Oe}$ leads also to further 
complication of the magnetization states in the 'after-switching' regime: the number of 
vortex-antivortex pairs which might exist simultaneously increases and the precession 
trajectory (limit cycle) of the average magnetization of the point-contact area becomes 
quasiperiodic. The complete analysis of the corresponding dynamics will be also presented 
elsewhere.

\begin{figure}[h]
\centering
{\includegraphics
[scale=0.8, bb=5cm -2cm 15cm 23cm]
{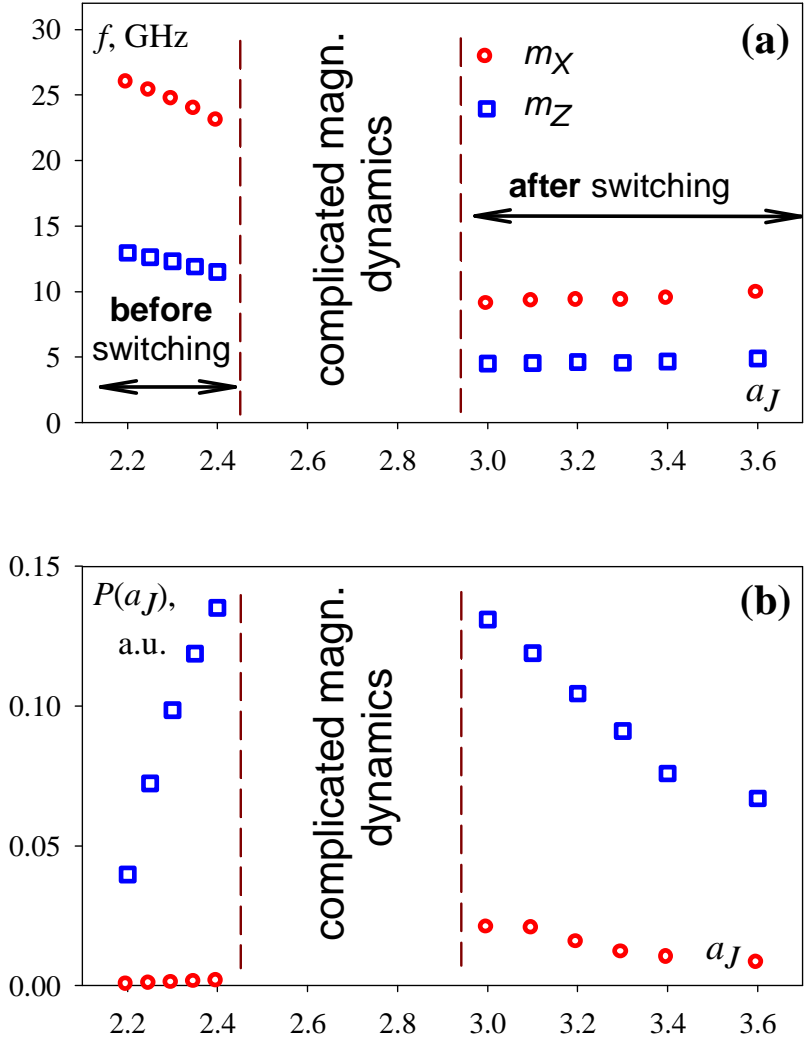}}
\caption
{
The same as in Fig. \ref{fig:TwoPrecRegimes_noOeField}, but for the magnetization dynamics
simulated including the current-induced magneic field. In the intermediate interval of
the spin torque amplitudes $a_J$ between two dashed lines the magnetization under the point
contact area demonstrates a very complicated dynamics, which can not be described as 
a steady-state precession with a simple closed limit cycle (see Fig. 
\ref{fig:ComplicatedMagnDyn_withFld}).
}
\label{fig:PrecRegimes_withOeField} 
\end{figure}

\begin{figure}[h]
\centering
{\includegraphics
[scale=0.8, bb=5cm -2cm 15cm 18cm]
{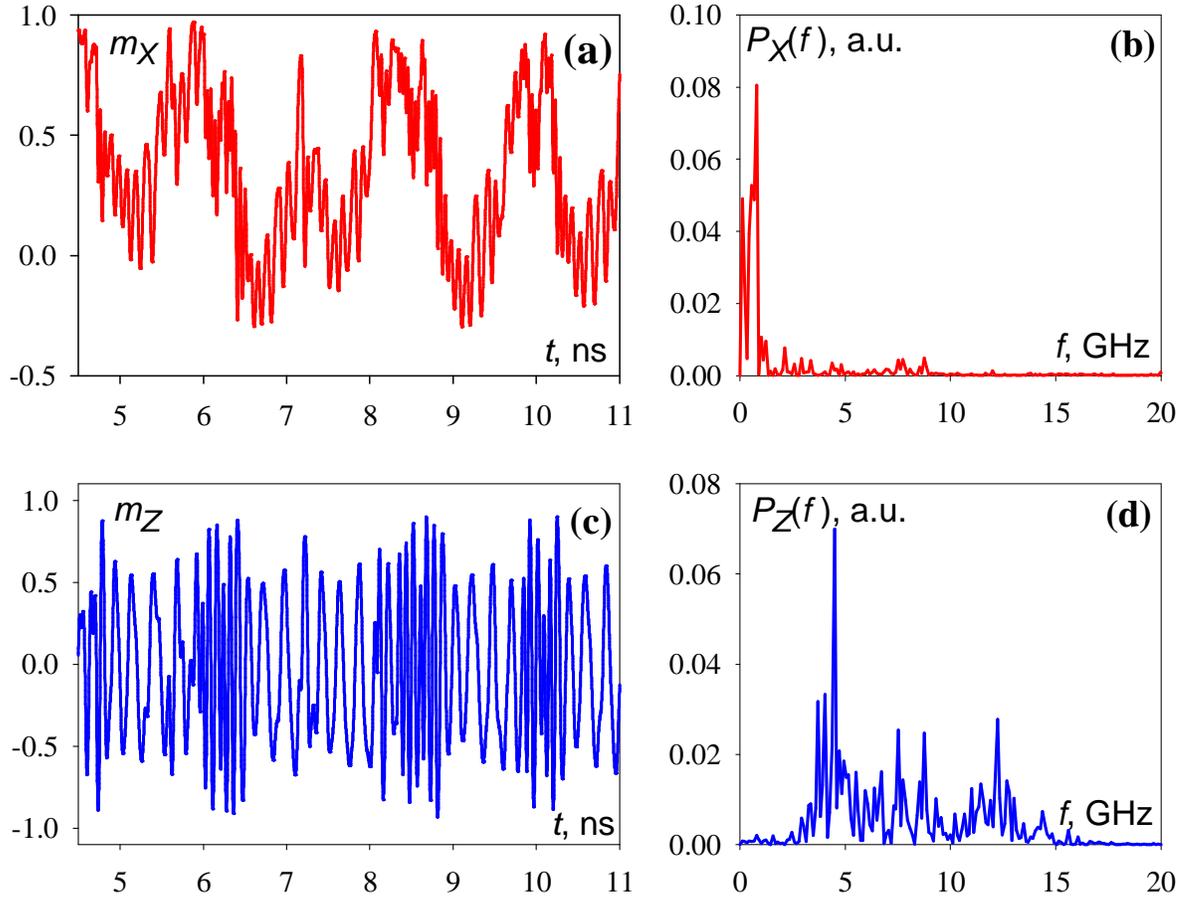}}
\caption
{
Typical time dependencies of $m_x^{\rm av}$ (a) and $m_z^{\rm av}$ (c) magnetization projections 
(averaged over the point contact area) at $a_J = 2.8$ when the Oersted field is taken into
account. Corresponding oscillation power spectra are shown in panels (b) and (d).
}
\label{fig:ComplicatedMagnDyn_withFld} 
\end{figure}

The major effect of the current-induced field is, however, not the quantitative changes
in the 'before-' and 'after-switching' precession modes discussed above, but the
appearance of a new intermediate regime in-between these two current regions. Corresponding
interval is marked in Fig. \ref{fig:PrecRegimes_withOeField} by the legend 'Complicated
magnetization dynamics'. For currents within this interval ($2.45 < a_J < 2.9$ for
parameters used in our simulations) the $x-$projection of the magnetization under the 
contact area exhibits relatively rare transitions between the values close to the maximal
possible value $m_x^{\rm max} = 1$ and values close to $m_x^{\rm av} = 0$ 
(Fig. \ref{fig:ComplicatedMagnDyn_withFld}a). Correspondingly, its power spectrum 
(Fig. \ref{fig:ComplicatedMagnDyn_withFld}b) has a large component at (relatively) low 
frequencies. The $z-$projection of the magnetization (in-plane projection perpendicular 
to the external field) oscillates with very different frequencies depending on the 
$m_x$-value, i.e., on the magnetization configuration in the point contact region: 
for nearly homogeneous magnetization state ($m_x^{\rm av}$ close to 1) the oscillation 
frequency of $m_z^{\rm av}$ is much higher than for a strongly inhomogeneous configuration 
(small values of $m_x^{\rm av}$). Oscillation power spectrum of $m_z^{\rm av}$ consists 
of several relatively broad lines, which quantitative analysis requires better 
simulation statistics (longer runs) than those which could be carried out up to now.

\section{Comparison with experimental data}
\label{sec:CompExpData}

In this last section we briefly compare our simulation data with experimental results
from Ref. \onlinecite{Rippard2004a}. 
First, we note that several important features of experimentally observed magnetization
precession in the point contact setup could be reproduced by our simulations. In 
particular, in the oscillation regime before switching we have obtained, in accordance with
Ref. \onlinecite{Rippard2004a}, very narrow spectral lines (in our simulations the
linewidth was $\Delta f < 50$ MHz) and nearly linear decay of the oscillation frequency
with increasing current (see Fig. \ref{fig:TwoPrecRegimes_noOeField}a and 
\ref{fig:PrecRegimes_withOeField}a). Such extremely small linewidths can be naturally 
explained by a smooth variation (in space) of the magnetization configuration, which is due
to the absence of strong demagnetizing field effects characteristic for nanoelements
in the columnar geometry (see, e.g., 
Ref. \onlinecite{Kiselev2003,Kiselev2004,SimulColumnGeom}). The decrease of the 
oscillation frequency with increasing current is a consequence of the growing precession
amplitude when the current (and hence - the spin torque magnitude) is increased, which
for non-linear oscillations results in the larger precession period.

\begin{figure}[h]
\centering
{\includegraphics
[scale=0.8, bb=5cm -2cm 15cm 18cm]
{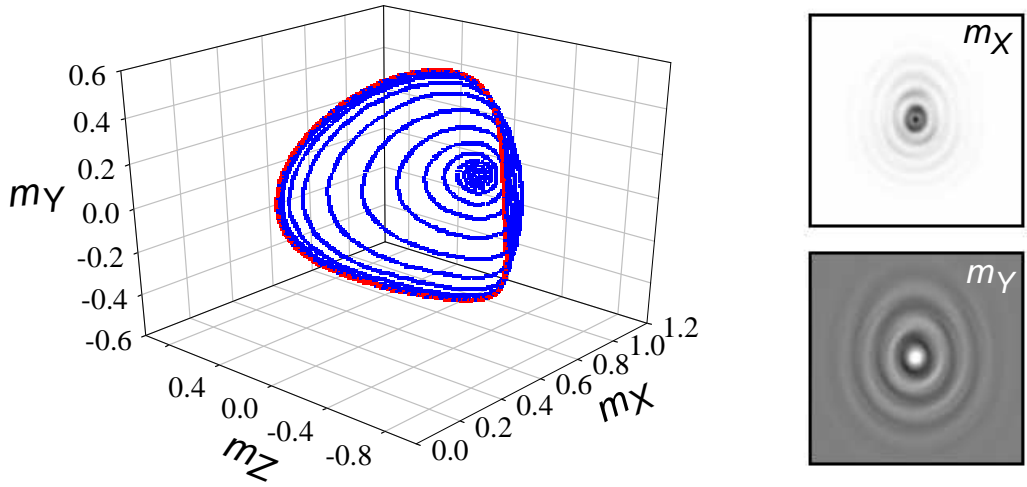}}
\caption
{
Steady-state precession {\it before} switching of the contact area: 3D trajectory of 
the magnetization ${\bf m}^{\rm av}$ averaged over the contact area (left panel) and 
snapshots of the waves emitted from the contact area shown as grey-scale maps of the component 
$m_x({\bf r})$ along the external field and $m_y({\bf r})$ - perpendicular to the layer plane 
(right panels). On the $m_y({\bf r})$-map the superposition of the {\it two} waves with the 
wavelengths corresponding to the precession frequencies of (i) longitudinal ($m_x$) and (ii) 
transverse ($m_y$ or $m_z$) magnetization components is clearly seen. Physical size of 
the images shown on the right is $900 x 900$ nm$^2$.
}
\label{fig:PrecBeforeSw_noOeField} 
\end{figure}

\begin{figure}[h]
\centering
{\includegraphics
[scale=0.8, bb=5cm -2cm 15cm 18cm]
{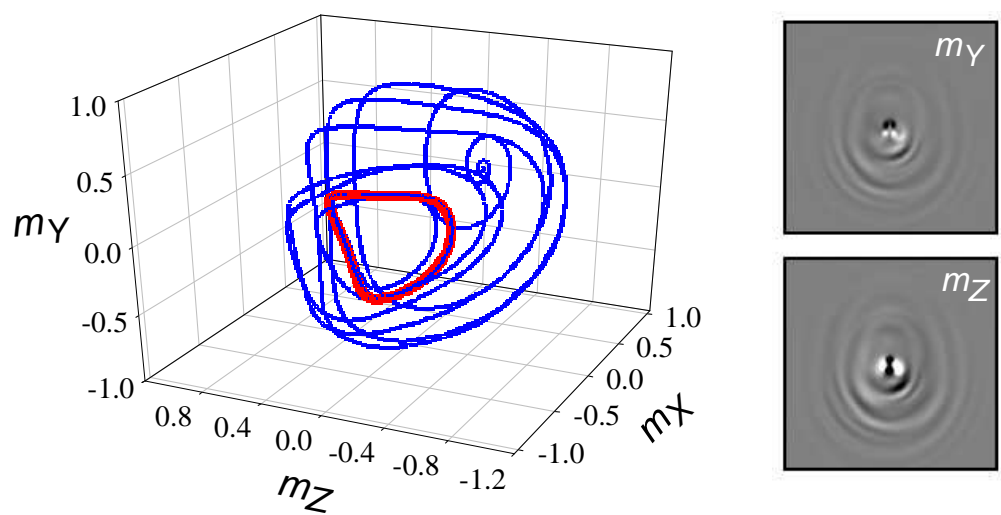}}
\caption
{
The same as in Fig. \ref{fig:PrecBeforeSw_noOeField} for the steady-state precession
{\it after} switching of the magnetization under the contact. Two compact wave packages
emitted from the contact area and their propagation direction are marked by white arrows.
}
\label{fig:PrecAfterSw_noOeField} 
\end{figure}

\begin{figure}[h]
\centering
{\includegraphics
[scale=0.8, bb=5cm -2cm 15cm 23cm]
{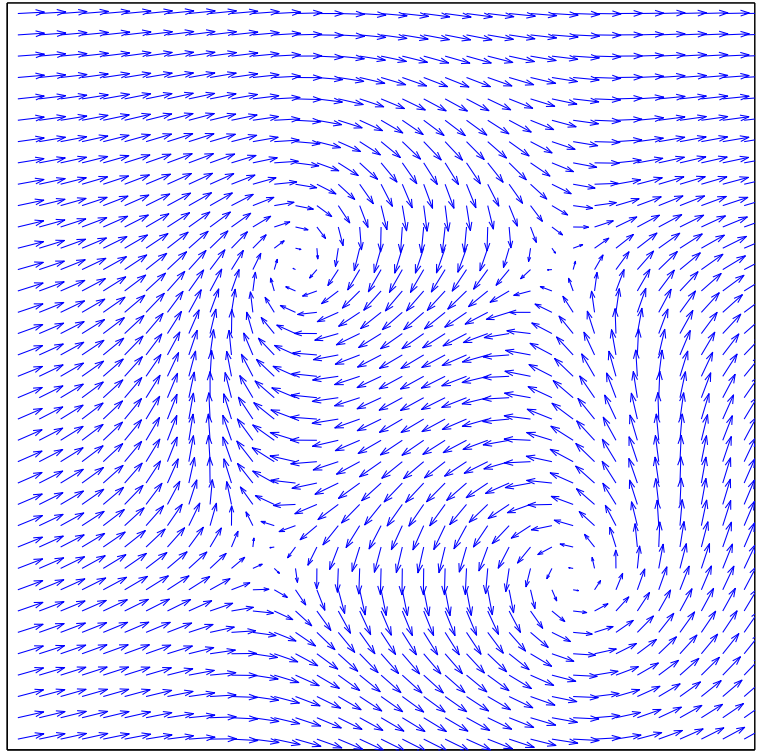}}
\caption
{
Magnetization configuration of the central part of the images presented in Fig.
\ref{fig:PrecAfterSw_noOeField} shown as arrows, which represent the orientation and
magnitude of the in-plane magnetization. The physical size corresponding to this arrow
map is $\approx 90 \times 90$ nm$^2$ (the point contact diameter is $D = 40$ nm).
}
\label{fig:ArrowMagnConf_noOeField} 
\end{figure}

More important, however, are the disagreements between simulated and measured data, among
which the {\it qualitative} one - the existence of at least two oscillation regimes for 
the simulated precession (whereby only one precession regime was observed experimentally) -
is the major problem. There are several possible reasons for this discrepancy, among
them (i) the simplicity of the torque term used in simulations (for more complicated forms
which should be tested next, see Ref. \onlinecite{Slonczewski2002,Xiao2004}), (ii) 
influence of the SPC-induced effective field recently measured in 
Ref. \onlinecite{Zimmler2004}, (iii) exchange weakening within magnetic layers resulting
from the local Joule heating (in the point contact geometry this effect
may be especially pronounced due to high current densities required to induce magnetization
oscillations) and (iv) substantial contribution of the layer regions outside the point contact 
area to the measured microwave oscillation spectra. In particular, for the last mentioned
reason, the signal from these regions would be present in the regime before switching, but 
nearly absent for the 'after-switching' mode (as it can be seen from the comparison of
right panels of Fig. \ref{fig:PrecBeforeSw_noOeField} and \ref{fig:PrecAfterSw_noOeField}), 
thus strongly enhancing the oscillation power in the regime before switching compared to the
second regime. However, to clarify whether it is really the case, calculations of the
current distribution for the concrete experimental setup are required.

Another interesting problem is the existence of a strong second harmonic in the experimentally
measured spectrum. Its presence could be caused by the local (under the point contact) 
deviation of the underlayer magnetization from the external field direction. This 
would lead to the contributions from the oscillations of both longitudinal ($m_x$) and 
transverse ($m_z$) in-plane magnetization components, thus providing the required
second harmonic ($m_x$) and basic ($m_z$) oscillation frequency \cite{BerkovEllipse2005}. Such
a magnetization deviation from the $H_0$ direction could exist due, e.g., to the random 
magnetic anisotropy of $Co_{90}Fe_{10}$ crystallites. Another explanation of the second 
harmonic presence could be the above mentioned contribution of the area around the point 
contact, because due to the local conservation of the magnetic moment magnitude the
waves of both $m_x$- and $m_z$-components contain both frequencies 
(see Fig. \ref{fig:PrecBeforeSw_noOeField}, right panels).

\section{Conclusion}

In conclusion we point out that full-scale micromagnetic simulations (performed in frames 
of the Slonczewski formalism) of the magnetization dynamics induced by a spin-polarized 
current in the point contact geometry recover several important features of experimental 
observations, like very narrow spectral lines and current dependence of the oscillation 
frequency. However, if we assume that the measured signal comes solely from the magnetization 
oscillation under the point contact area, simulations results exhibit serious qualitative 
disagreements with experimental data, the main of which is the existence of at least two 
precession modes in the two current intervals corresponding to (i) the precession around 
the external field direction ('before-switching' mode) and (ii) around the direction 
opposite to $H_0$ ('after-switching' mode). This disagreement clearly shows that further 
refinement of theoretical models is required for understanding of the spin torque induced
magnetization excitations in point-contact experiments.

\newpage

\end{document}